\documentclass[preprint]{aastex}
\usepackage{graphicx, float}
\usepackage{epstopdf}
\usepackage{subcaption}
\usepackage{natbib}
\usepackage{mathtools}
\usepackage{lineno}
\usepackage{booktabs}
\makeatletter
\let\@dates\relax
\makeatother
\begin{document}

\title{The Scaling of the RMS with Dwell Time in NANOGrav Pulsars}

\author{Emma Handzo\altaffilmark{1}\altaffiltext{1}{now at the University of Texas Rio Grande Valley}, B. Christy, Andrea N. Lommen, Delphine Perrodin\altaffilmark{2}\altaffiltext{2}{now at INAF - Osservatorio Astronomico di Cagliari, Cagliari, Italy}}
\affil{Department of Physics and Astronomy, Franklin and Marshall College\\
415 Harrisburg Pike, Lancaster, Pennsylvania, 17603}

\begin{abstract}
Pulsar Timing Arrays (PTAs) are collections of well-timed millisecond pulsars that are
being used as detectors of gravitational waves (GWs). 
Given current sensitivity, projected improvements in PTAs 
and the predicted strength of the GW signals, the detection of 
GWs with PTAs could occur within the next decade.  
One way we can improve a PTA is to reduce the measurement noise 
present in the pulsar timing residuals.
If the pulsars included in the array display uncorrelated noise, 
the root mean square (RMS) of the timing residuals
is predicted to scale as $\mathrm{T}^{-1/2}$, where T is the dwell time per observation.
In this case, the sensitivity of the array can be increased by
increasing T. We studied the 17
pulsars in the five year North American Nanohertz Observatory for
Gravitational Waves (NANOGrav) data set to determine if the noise in the timing residuals of the pulsars
observed was consistent with this property. For comparison, we
performed the same analysis on PSR B1937+21, a pulsar that is known 
to display red noise. With this method, we find that 15 of the 17 NANOGrav 
pulsars have timing residuals consistent with the inverse square law. The data also suggest that 
these 15 pulsars can be observed for up to eight times as long while still exhibiting an RMS that scales as root T.
\end{abstract}

\section{Introduction}

A Pulsar Timing Array (PTA) is a collection of millisecond pulsars
monitored frequently for imprints of gravitational waves (GWs).
Currently there are three PTA collaborations:
the European Pulsar Timing Array (EPTA) \citep{Ferdman10}, the Parkes
Pulsar Timing Array (PPTA) \citep{Manchester13}, and the North
American Nanohertz Observatory for Gravitational Waves (NANOGrav)
\citep{Jenet09}. In partnership with each other, they make up the
International Pulsar Timing Array (IPTA) \citep{Hobbs10}.  Using a PTA
as an instrument for GW detection, these collaborations search for correlated
disturbances in the pulse arrival times. 

\cite{Siemens13} estimate that detection of GWs will happen
in the next decade, with uncertainties arising from unknowns such as the amplitude
of the GW signal, the number of pulsars added to the array in the future, and the noise
level of the pulsar themselves.  The appraisal of pulsar noise is key in estimating
the sensitivity of the experiment,
and in this manuscript
we employ a simple test to provide some knowledge of the noise behavior in
the NANOGrav pulsars. We are concerned specifically with the root mean square (RMS) residual
as an estimator for the noise amplitude.
We summarize briefly here the expected scaling of
the sensitivity of the overall experiment with RMS.
\cite{Siemens13} calculates the
average GW signal-to-noise (GW S/N) ratio for a PTA that samples the
pulsars at regular intervals, assuming each of these pulsars has the
same uncorrelated noise level. They show that in the weak-signal limit
(where the uncorrelated noise dominates the GW signal), the GW S/N is
inversely proportional to the square of the RMS. In the intermediate
regime (where the uncorrelated noise is roughly equivalent with the GW
signal), the GW S/N has a more complicated relationship to the RMS,
but the sensitivity to GWs still increases with
decreasing RMS. Finally, in the strong-signal limit (where the GW
signal dominates the noise), the GW S/N is determined by the cadence
of the observations. 
Reducing the RMS residual is therefore a crucial PTA strategy
 until we find ourselves in the strong-signal regime.

Although there are multiple avenues for noise (and therefore RMS) reduction,
this paper focuses on the possibility of decreasing the RMS by 
increasing the {\em dwell} time,
the amount of observation time spent on a pulsar to obtain a single arrival time.
We roughly expect a pulsar's timing residuals to be Gaussian
distributed and, consequently, we expect the RMS of
the residuals to scale as the inverse square root of the 
dwell time.  Thus we commonly assume that we can improve the
sensitivity of the experiment simply by increasing the dwell time
\citep{Jenet05,Jenet06}.
 
The presence of red noise changes the situation.  There are many
possible sources of the red noise: measurement noise, GWs
\citep{Lommen01}, intrinsic noise \citep{Hobbs06}, objects orbiting
the pulsar \citep{Hobbs06,Shannon13}, or interstellar medium (ISM)
effects \citep{Hemberger08}.  Red noise likely exists at some level in
all pulsars, and it becomes more important as we reduce the white
noise.  When red noise begins to dominate the overall noise, we can no
longer expect the RMS to decrease with increasing dwell time
\citep{Jenet06}.

The goal of this paper is to determine the extent to which increasing
the dwell time on NANOGrav pulsars will reduce the
noise in the timing residuals. To test our ability to decrease the
RMS, we cannot go back and increase the amount of dwell time for each
pulsar in past observations, but we can mimic this by adding up
adjacent observations to simulate longer dwell times. Throughout the
paper, we use ``N'', the number of adjacent observations being added
together, which is a surrogate to ``T'', the dwell time. This allows
us to discuss the ``root N'' behavior of the pulsars, which should be
read analogous to ``root T''. This is a common test done in pulsar
timing \citep{Helfand75,Cordes81,Liu12,Dolch14}. This paper is
organized as follows: in \S2, we demonstrate the expectation of the root N
dependence of the RMS residual, highlighting the conditions that must
be met. In \S3, we describe how we applied this to the NANOGrav data
and present the results. In \S4, we discuss whether the pulsars display 
this dependence and the limits to the test resulting from the timing model fit. 
In \S5, we summarize our findings, discuss our results in the context of
 EPTA and PPTA data, and suggest how future work will help us understand 
 the situation better..

\section{The N$^{-1/2}$ Test}
\label{sec:weighted_rms} 
 
A timing model for a pulsar is generated after a pulsar has been
observed for a long period of time. It predicts when the next pulse
from the pulsar should arrive \citep{Hobbs06}. The timing model
yields a timing residual, which is the measured pulse arrival time
minus the predicted pulse arrival time:
\begin{equation}
t=t_{\mathrm{measured}}-t_{\mathrm{predicted}}
\end{equation}
We typically use the RMS of these residuals as a measure to characterize the total noise in the dataset.
A lower RMS residual is indicative of residuals that are close to zero.
While many pulsars follow the trend
$\mathrm{RMS}\propto\mathrm{N}^{-1/2}$, where N is the number of observations made,
 not all pulsars do. We need to
determine which pulsars' RMS residual decreases with increasing N
and which do not, so we can make informed choices about which pulsars
should be included in the PTA and for how long they should be observed.

We want to determine the effect of increasing the dwell time on the RMS of a pulsar.  
As a proxy, we make this comparison by averaging N successive points together.  
Then, we study how the RMS of the timing residuals changes as a function of N.

First, let us demonstrate the existence of the root N relation and the conditions under which we expect it to hold true. We average an original data set $\mathbf{T}$
by every N points. This gives us a new data set, $\mathbf{T^\prime} = \{t^\prime_1 ,t^\prime_2 ... t^\prime_M\}$, 
where M is the number of points in the new, averaged dataset.  If Q is 
the total number of points in the original dataset, then Q=M$\times$N. 
To create $\mathbf{T^\prime}$ we first create sets of samples of $\mathbf{T}$ of every $N$th number:
$\mathbf{T}_1=\{t_1,t_{N+1},....\}$,
$\mathbf{T}_2=\{t_2,t_{N+2},.....\}$ and so on. The new points are
calculated using
${\mathbf{T}^\prime}=\{\frac{t_1+t_2+...+t_N}{N},\frac{t_{N+1}+t_{N+2+...+t_{2N}}}{N},...\}=\frac{\mathbf{T}_1+\mathbf{T}_2+...+\mathbf{T}_N}{N}$. The relationship between
$\mathrm{Var}({\mathbf{T'}})$ and $\mathrm{Var}(\mathbf{T})$, becomes:
\begin{eqnarray}
\mathrm{Var}({\mathbf{T^\prime}})=\frac{\mathrm{Var}(\mathbf{T})}{N}
\end{eqnarray}
It follows trivially that:
\begin{equation}
\mathrm{RMS}_{{\mathbf{T'}}} =\frac{1}{\sqrt{N}}\mathrm{RMS}_{\mathbf{T}}
\label{eqn:rms}
\end{equation}

It is the relationship contained in Equation \ref{eqn:rms} that we investigate
in this paper. This relationship has been studied by many other authors (such 
as \citet{Helfand75, Cordes81, Jenet05, Liu12}), whose results were similar to ours. 
Note that the veracity of the relation depends upon the residuals
being statistically independent. If this is not the case, e.g. in red noise,
$\mathrm{Var}(P_1+P_2)\ne\mathrm{Var}(P_1)+\mathrm{Var}(P_2)$, and Equation 2 is not valid. Also, if the distribution is non-stationary, 
i.e. it is changing in time, then the relationship $\mathrm{Var}(P_1)\ne\mathrm{Var}(P_2)$ 
would be true, and we would not expect the RMS of the pulsars to decline as $\mathrm{N}^{-1/2}$.
Recent works by \citet{Ellis12} and references 
therein state that the NANOGrav data appears to display noise patterns that allow us to assume stationary statistics
with this data set.

In PTA data sets, each data point is assigned an uncertainty, $\sigma$, based on the 
S/N of the data taken. The uncertainty can vary due to a variety of conditions, 
such as scintillation from the ISM. To account for this variation, our averaging was performed 
by weighting the points with $1/\sigma^2$. We performed a simulated test to verify 
if residuals with unequal errors still follow the root N trend.  We took the J1713+0747 
dataset and replaced the data points with values drawn from a Gaussian distribution. 
The unequal uncertainties from the original dataset were kept. 
We performed the test on this dataset and found that the RMS of the simulated data follows the expected
root N relationship even with the unequal errors.

\section{Analysis}
\label{sec:analysis}
In this paper, we analyze the residuals of the timing data set on the 17 
pulsars presented by \citet{Demorest13} in order to determine 
whether the NANOGrav pulsars' timing noise follows the root N relationship. Currently the NANOGrav 
pulsars are allotted roughly equal observing time of 15-45 minutes
within NANOGrav's total telescope time. The 17 pulsars that were examined in this paper
 were observed from 2005 to 2010 at either the Arecibo 
Observatory or the Green Bank Telescope, in frequencies ranging from 330MHz to 2500MHz. The one exception is 
PSR J1713+0747, which was observed at both telescopes.

We demonstrated in the previous section that we expect the RMS residual to be proportional to
$\mathrm{N}^{-1/2}$ for pulsars whose residuals display uncorrelated noise when calculating the
RMS of a residual data set $\mathbf{T}= {t_1, t_2, ....t_Q}$. For our data set, N is the 
number of data points that are being averaged together. Since each data point is 
roughly the same observation time, N is directly proportional to the total dwell time, e.g. if you average
every two points together, the resulting point is the expected timing residual of that pulsar 
being observed for twice as long. As discussed in \S2, these data have unequal 
uncertainties so we weight them by $1/\sigma^2$ to account 
for the fact that some observations are 
not as accurate as others, due to the observing conditions. 

In the NANOGrav observing program that \citet{Demorest13} uses,
data from an observing epoch are split into 4MHz sub-bands in order 
to study the frequency evolution of the profile in greater detail. As we are interested 
in the behavior over long time scales \citep{Wackerly}, we begin by averaging over these frequency sub-bands as follows:

\begin{equation}
t_i = \frac{\displaystyle\sum\limits_{j=1}^{M_i}({t_{i,j}/\sigma_{i,j}^2})}{\displaystyle\sum\limits_{j=1}^{M_i}(1/\sigma_{i,j}^2)}
\end{equation}
where $t_{i,j}$ is the $j$th residual on day $i$, with $\sigma_{i,j}$ representing the TOA error, resulting from the cross-correlation with a template (see \citet{Demorest13}), and $M_i$ is the total number of residual points on this day.
Uncertainties are given by the standard deviation:
\begin{equation}
\sigma_{i} = \sqrt{\frac{\displaystyle\sum\limits_{j=1}^{M_i}((t_{i,j} - \overline{t})^2/\sigma_{i,j}^2)}{(M_i-1) \displaystyle\sum\limits_{j=1}^{M_i}(1/\sigma_{i,j}^2)}}
\end{equation}
where $\overline{t}$ is the weighted mean of each data set \citep{Wackerly}. With the large 
number of points in a given epoch, typically $\sim$15, we expect the fit to the average to 
follow a $\chi^2$ distribution with $M-1$ degrees of freedom, and we weight the uncertainties
 by the reduced $\chi^2$. These calculations yield a 
reduced new data set, ($\mathbf{T}$) that
depicts one timing residual ($t_i$) and the standard deviation of the timing residual
($\sigma_i$) for each observing day $i$.
We use these reduced data sets to inspect how the RMS 
residual changes if we observe each pulsar for longer periods of time, or
in other words, as we increase N.

To begin, we calculate the weighted RMS (WRMS) of the
epoch-averaged residuals, $\mathbf{T} = \{t_1, t_2, ...t_Q\}$:
\begin{eqnarray}
\mathrm{WRMS}=\sqrt{\frac{\displaystyle\sum_{i=1}^Q((t_i-\overline{{t}})^2/(\sigma_i)^2)}{\displaystyle\sum_{i=1}^Q(1/(\sigma_i)^2)}}.
\label{eqn:wrms}
\end{eqnarray}
The uncertainty in this residual WRMS is given by:
\begin{eqnarray}
\sigma_{\mathrm{WRMS}}=\sqrt{\frac{1}{\displaystyle\sum_{i=1}^Q{(1/(\sigma_i)^2)}}}
\label{eqn:sigmarms}
\end{eqnarray}
Next, we average every N residual data points that are consecutive in time to make a new data set $\mathbf{T'}$ just
as described above in Equation 4. The error for these points is found using error propagation and is similar to Equation 5. Note that, 
unlike when we calculated the daily averaging, we do not weight the 
uncertainty by $\chi^2$ since the number of points is small and the $\chi^2$ distribution 
is not well behaved. We let N have the range $\{1,B\}$ where B is equal to the number of residuals in the original set divided by the number of points being averaged together.
Again, the residual WRMS and $\sigma_{{\rm WRMS}}$
are calculated using Equations \ref{eqn:wrms} and \ref{eqn:sigmarms}.

We examine the rate at which the WRMS of the timing residuals for each of the
NANOGrav pulsars decline with respect to N.
The majority of the NANOGrav pulsars were observed in both a 
low-frequency (generally 800 MHz) and a high-frequency (generally 1400 MHz) band. The exceptions to this are PSR
J1713+0747, which was observed in the high and low bands at both the Arecibo Observatory (AO) 
and the Green Bank Telescope (GBT), and PSRs B1953+29,
J1853+1308 and J1910+1256, which were originally part of another
project so we have only the 1410 MHz observations for most of the
observation time \citep{Demorest13, Gonzalez11}. 

To quantify the extent each pulsar follows the root N behavior, we find the least squares fit of the data to WRMS=aN$^{-1/2}$, where a is the only fitted parameter. For each fit, a reduced $\chi^2$ was found using:
\begin{equation}
\mathrm{Reduced}~\chi^2=\frac{\displaystyle\sum_i((\mathrm{WRMS}_i-\mathrm{fit})^2/\sigma_i^2)}{\mathrm{\mathrm{K}_{\mathrm{dof}}}}
\end{equation}
where K$_{\mathrm{dof}}$ is the number of degrees of freedom found with K$_{\mathrm{dof}}=\mathrm{K}_{\mathrm{points}}-\mathrm{K}_{\mathrm{parameters}}-1=\mathrm{K}_{\mathrm{points}}-2$.

\begin{table}[htbp]
  \centering
  \title{$\chi^2$ values}
  \caption{Table 1 shows the goodness of the fit between the expected RMS value fitted to a slope of 
  $\mathrm{N}^{-1/2}$. The goodness of the 
  data is measured by the $\chi^2$ and reduced $\chi^2$.}
    \begin{tabular}{cccccc}
    \toprule
    \multicolumn{1}{c}{Pulsar} & Frequency (MHz) & Observatory & $\chi^2$ & Reduced $\chi^2$ \\
    \midrule
    \multicolumn{1}{c}{J0030+0451} & 1410 & AO & 4.46914295786 & 0.558642869733 \\
    \multicolumn{1}{c}{J0613-0200} & 1390 & GBT & 12.0463663442 & 0.708609784953 \\
    \multicolumn{1}{c}{J1012+5307} & 1390 & GBT & 14.2027958984 & 0.835458582261 \\
    \multicolumn{1}{c}{J1455-3330} & 1390 & GBT & 14.0405246474 & 0.825913214552 \\
    \multicolumn{1}{c}{J1600-3053} & 1400 & GBT & 2.11653100526 & 0.211653100526 \\
    \multicolumn{1}{c}{J1640+2224} & 1400 & AO & 84.0567098794 & 8.40567098794 \\
    \multicolumn{1}{c}{J1643-1224} & 1390 & GBT & 80.2630069467 & 4.72135334981 \\
    \multicolumn{1}{c}{J1713+0747} & 1400 & GBT & 4.54006173273 & 0.267062454867 \\
    \multicolumn{1}{c}{J1713+0747} & 2500 & AO & 10.9673468661 & 0.913945572174 \\
    \multicolumn{1}{c}{J1744-1134} & 1380 & GBT & 28.6140639164 & 1.68318023038 \\
    \multicolumn{1}{c}{J1853+1308} & 1410 & AO & 8.23077276824 & 0.54871818455 \\
    \multicolumn{1}{c}{B1855+09} & 1410 & AO & 13.7537378327 & 1.05797983329 \\
    \multicolumn{1}{c}{J1909-3744} & 1400 & GBT & 19.2218747739 & 1.37299105528 \\
    \multicolumn{1}{c}{J1910+1256} & 1410 & AO & 40.98978419 & 3.15306032231 \\
    \multicolumn{1}{c}{J1918-0642} & 1400 & GBT & 16.0827033438 & 0.946041373163 \\
    \multicolumn{1}{c}{B1937+21} & 1405 & AO & 271657.658795 & 15979.8622821 \\
    \multicolumn{1}{c}{B1953+29} & 1410 & AO & 10.0596679919 & 1.11774088799 \\
    \multicolumn{1}{c}{J2145-0750} & 1400 & GBT & 10.267139863 & 1.14079331812 \\
    \multicolumn{1}{c}{J2317+1439} & 430 & AO & 9.20820969886 & 0.708323822989 \\
    \bottomrule
   \end{tabular}
 \label{chi}
\end{table}

\begin{figure}[H]
        \centering
        \begin{subfigure}[b]{0.49\textwidth}
                \centering
                \includegraphics[width=\textwidth]{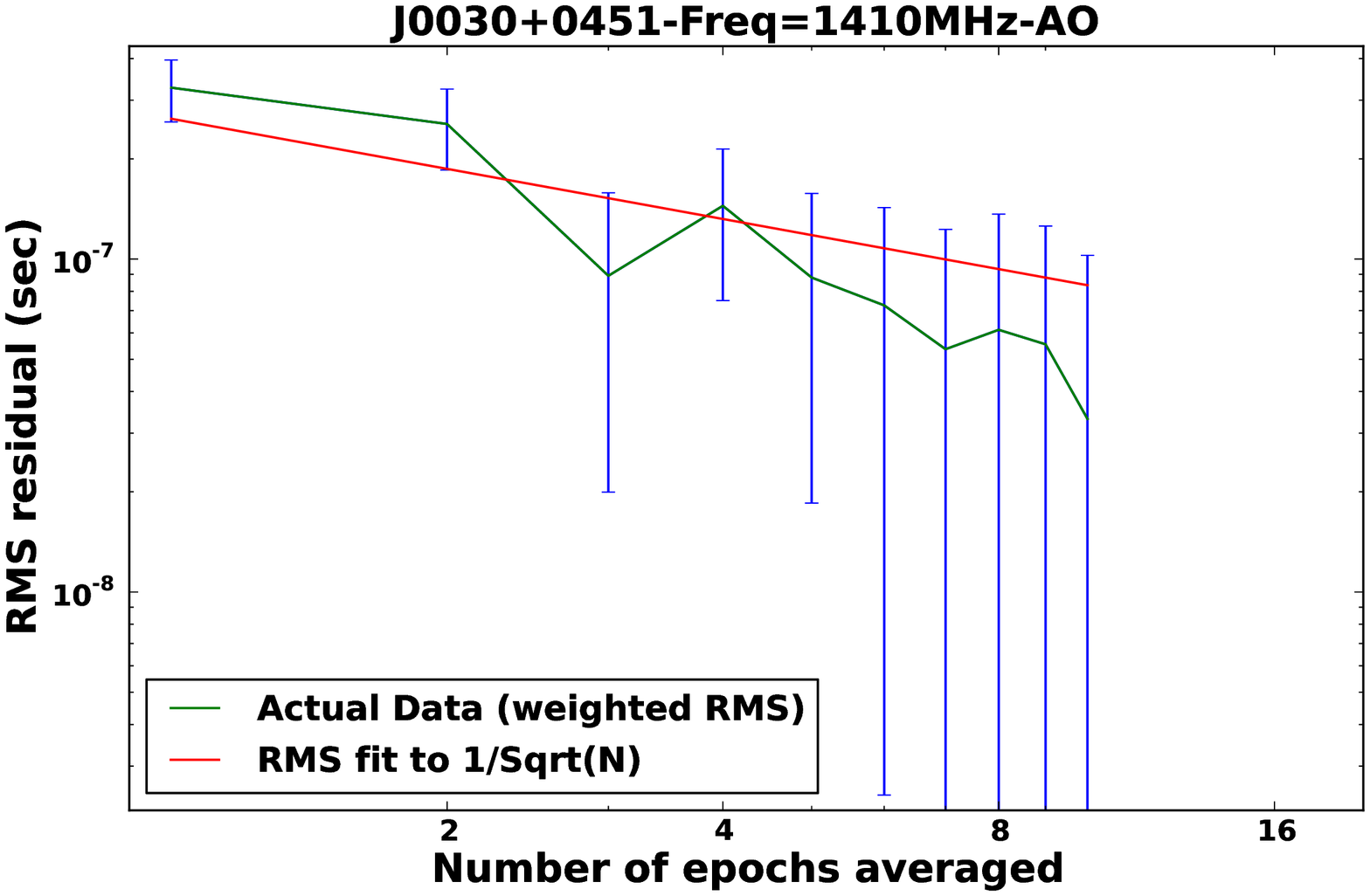}
                \caption{}
                \label{fig:residJ0030+0451b}
        \end{subfigure}%
        \begin{subfigure}[b]{0.49\textwidth}
                \centering
                \includegraphics[width=\textwidth]{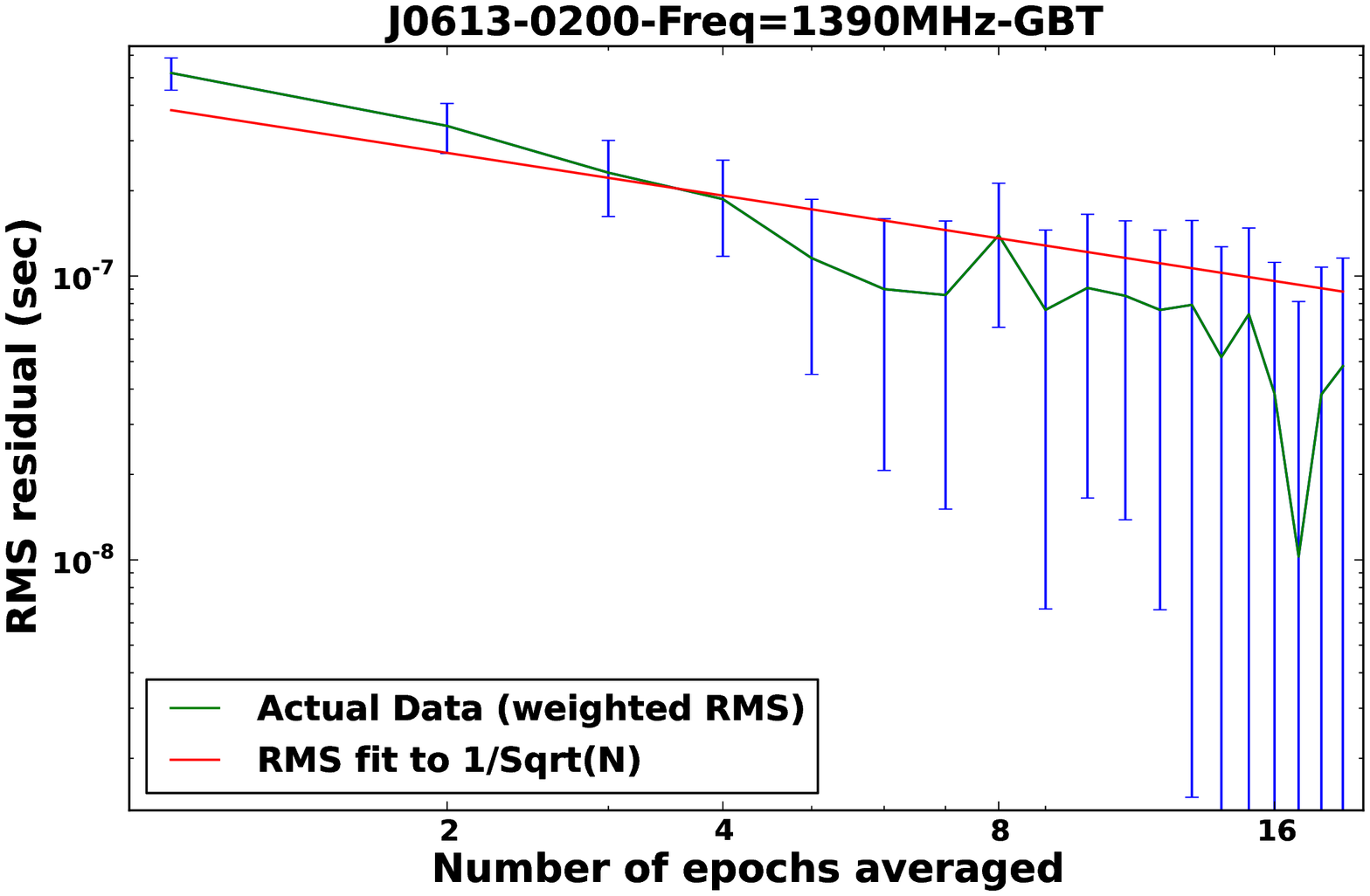}
                \caption{}
                \label{fig:residJ0613-0200b}
        \end{subfigure}
        \begin{subfigure}[b]{0.49\textwidth}
                \centering
                \includegraphics[width=\textwidth]{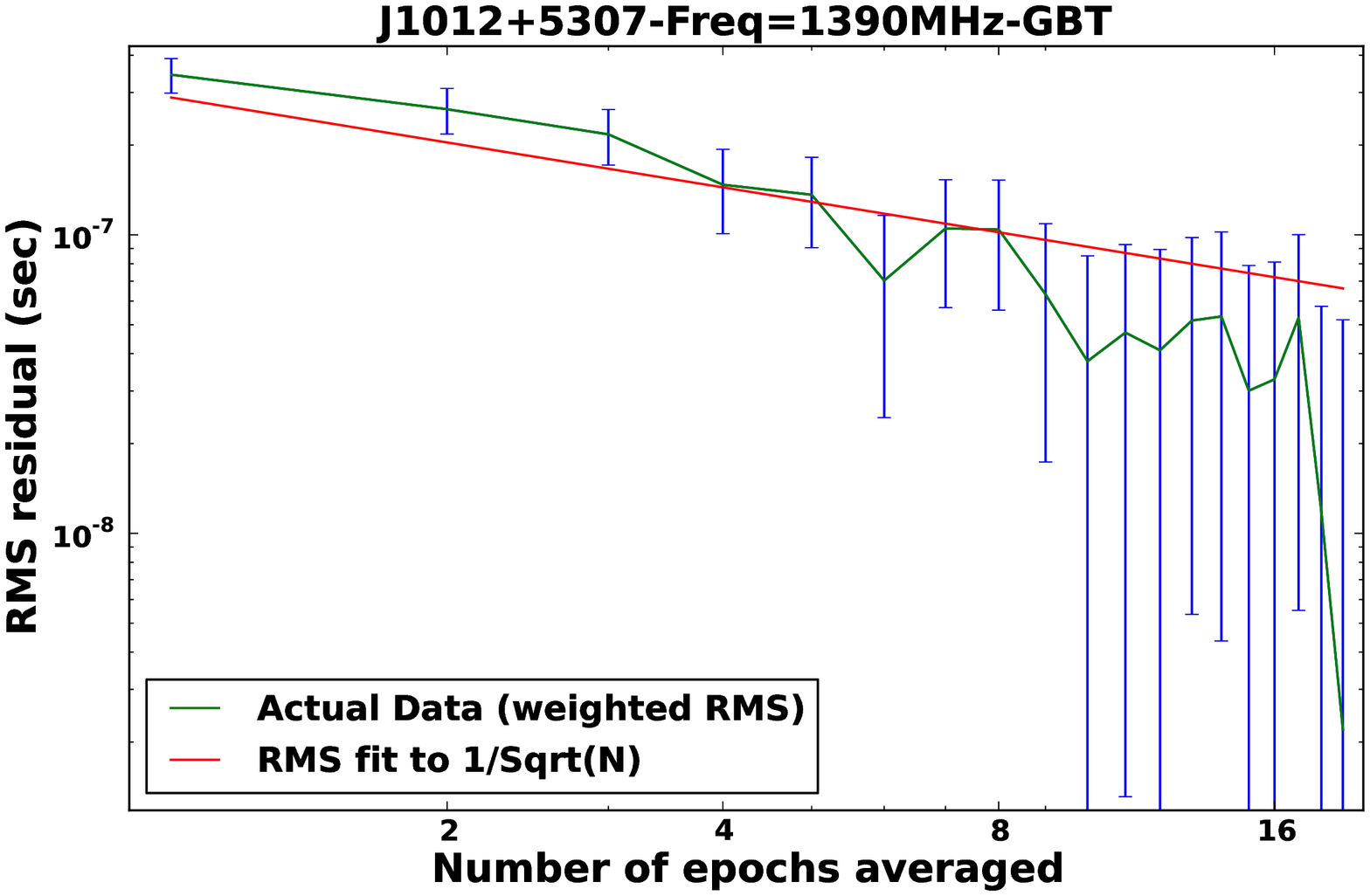}
                \caption{}
                \label{fig:residJ1012+5307b}
        \end{subfigure}%
        \begin{subfigure}[b]{0.49\textwidth}
                \centering
                \includegraphics[width=\textwidth]{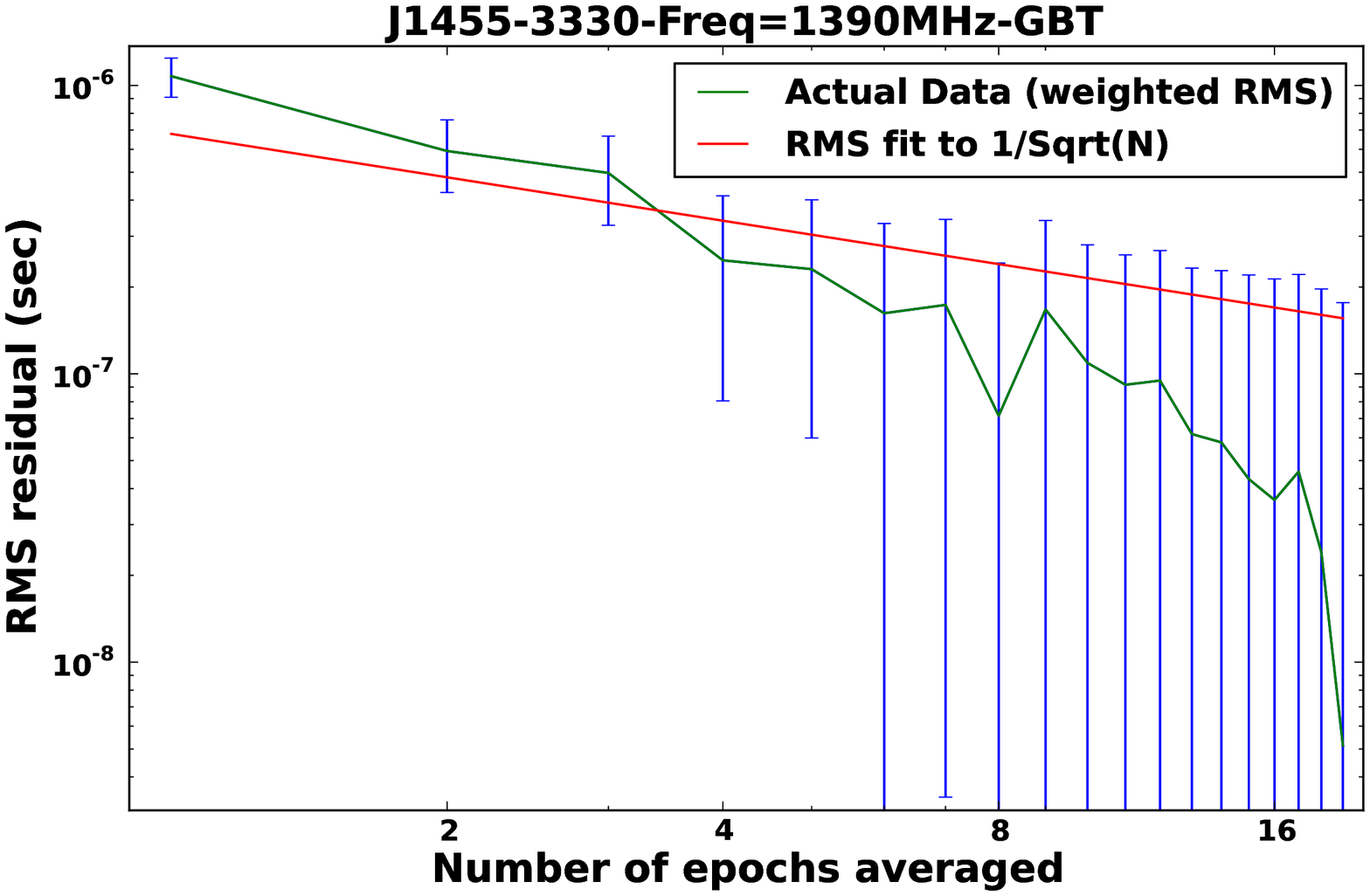}
                \caption{}
                \label{fig:residJ1455-3330b}
        \end{subfigure}
        \begin{subfigure}[b]{0.49\textwidth}
                \centering
                \includegraphics[width=\textwidth]{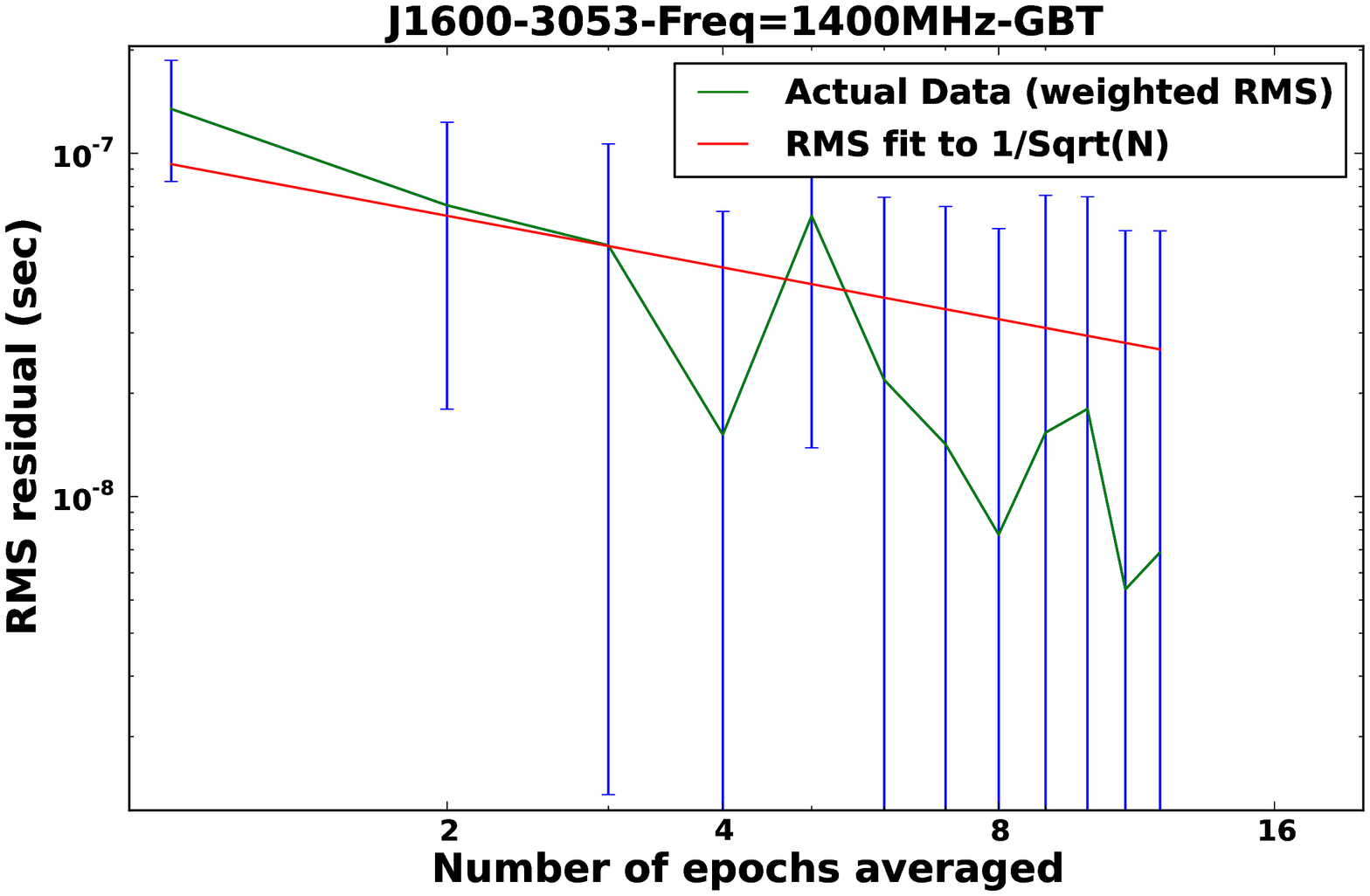}
                \caption{}
                \label{fig:residJ1600-3053b}
        \end{subfigure}%
        \begin{subfigure}[b]{0.49\textwidth}
                \centering
                \includegraphics[width=\textwidth]{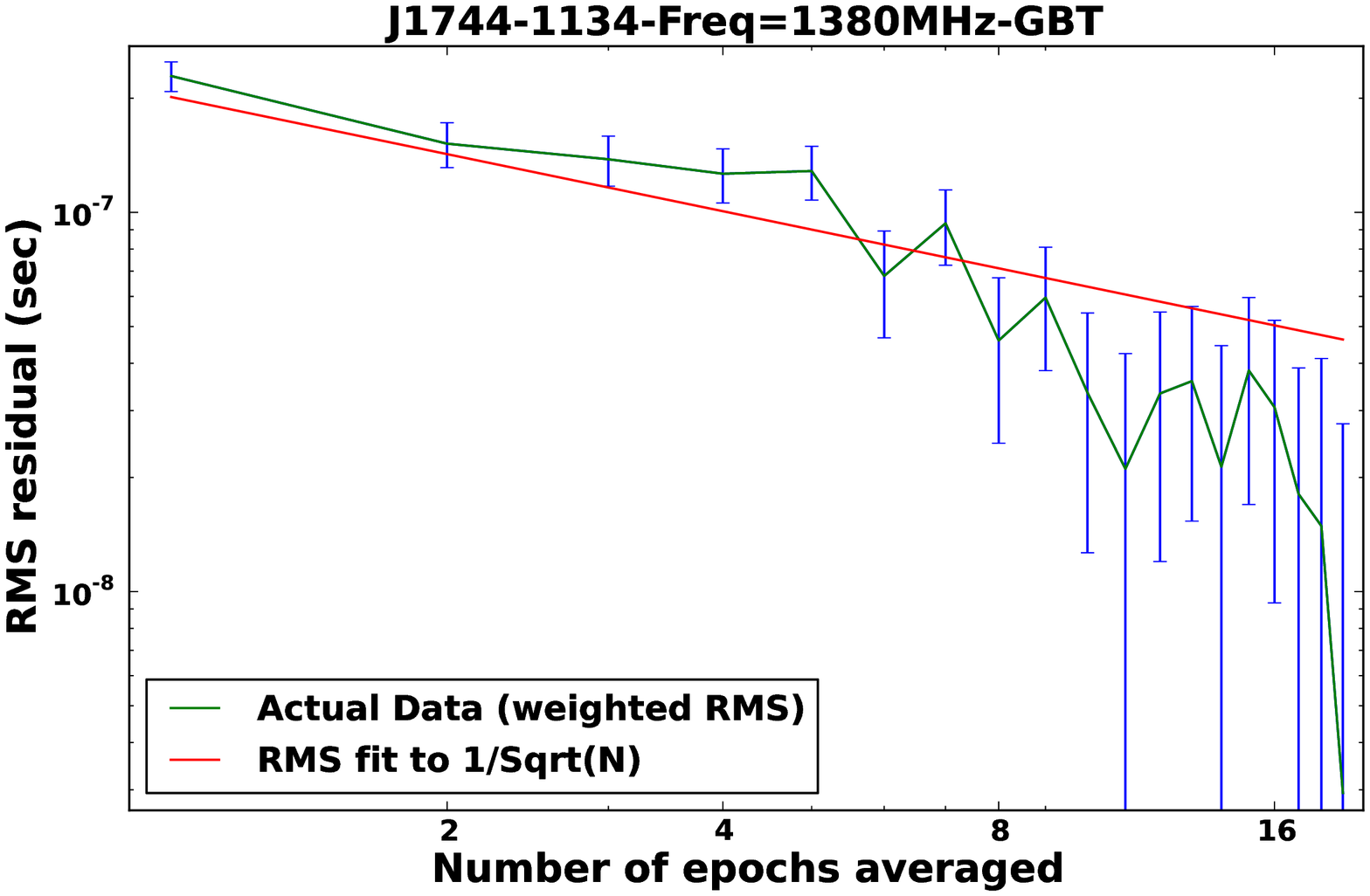}
                \caption{}
                \label{fig:residJ1744-1134b}
        \end{subfigure}
        \caption{RMS residual vs. the number of residuals averaged for the PSR J0030+0451 
        when observed at 1410 MHz (Panel a), PSR J0613$-$0200 at 1410 MHz 
        (Panel b), PSR J1012+5307 at 1390 MHz (Panel c), PSR J1455$-$3330 at 1390 MHz (Panel d), PSR      	J1600$-$3053 at 1400 MHz (Panel e), and PSR J1744$-$1134 at 1380 MHz (Panel f).
        The solid line represents the RMS fit to
          $\mathrm{N}^{-1/2}$.
          The plots show that within uncertainties, the
          RMS residual is proportional to $\mathrm{N}^{-1/2}$ for these
          pulsars.\label{fig:J0030+0451J0613-0200J1012+5307J1455-3330J1600-3053J17441134} }
\end{figure}

\begin{figure}[H]
        \centering
        \begin{subfigure}[b]{0.49\textwidth}
                \centering
                \includegraphics[width=\textwidth]{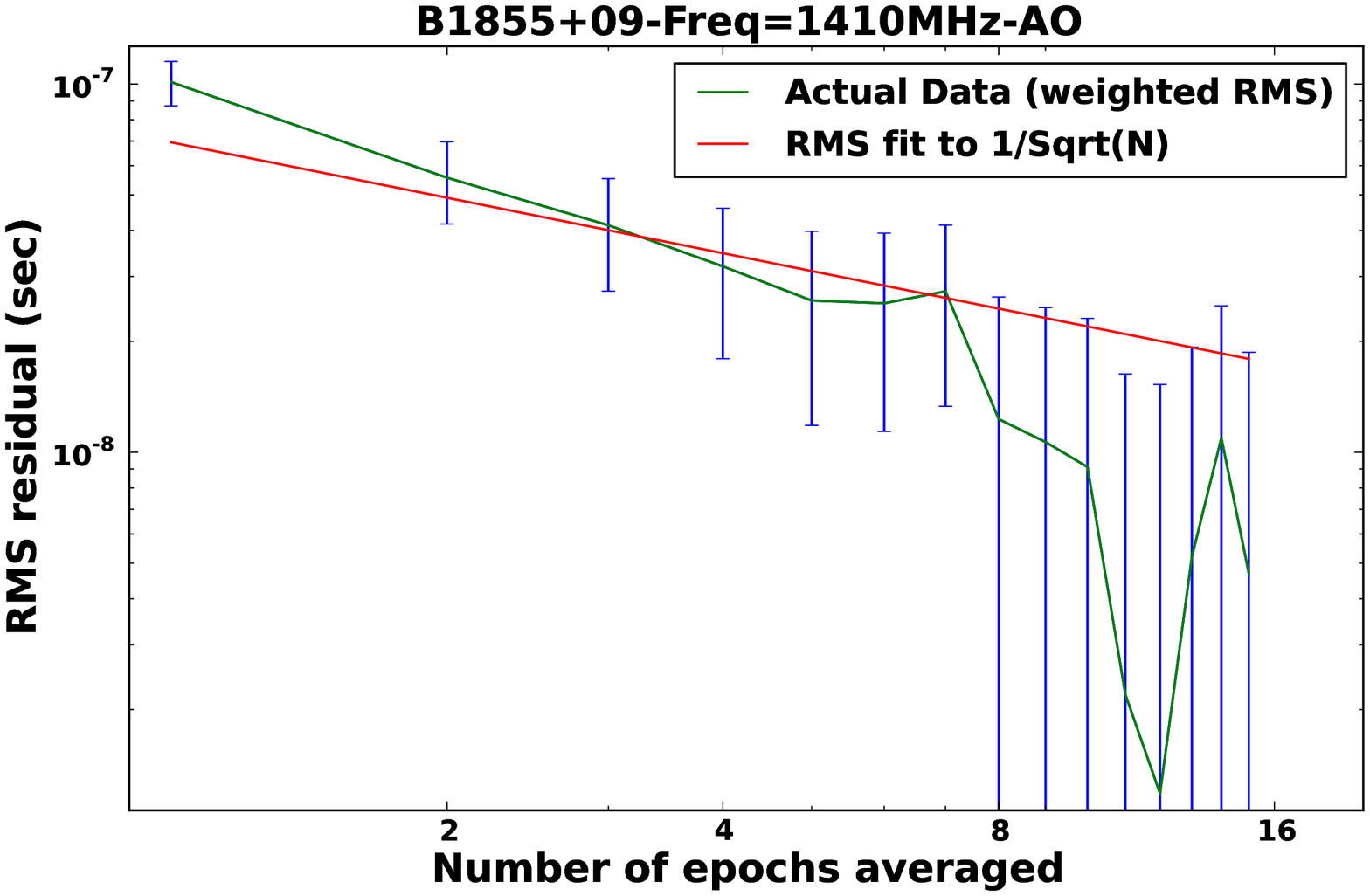}
                \caption{}
                \label{fig:residB1855+09b}
        \end{subfigure}%
        \begin{subfigure}[b]{0.49\textwidth}
                \centering
                \includegraphics[width=\textwidth]{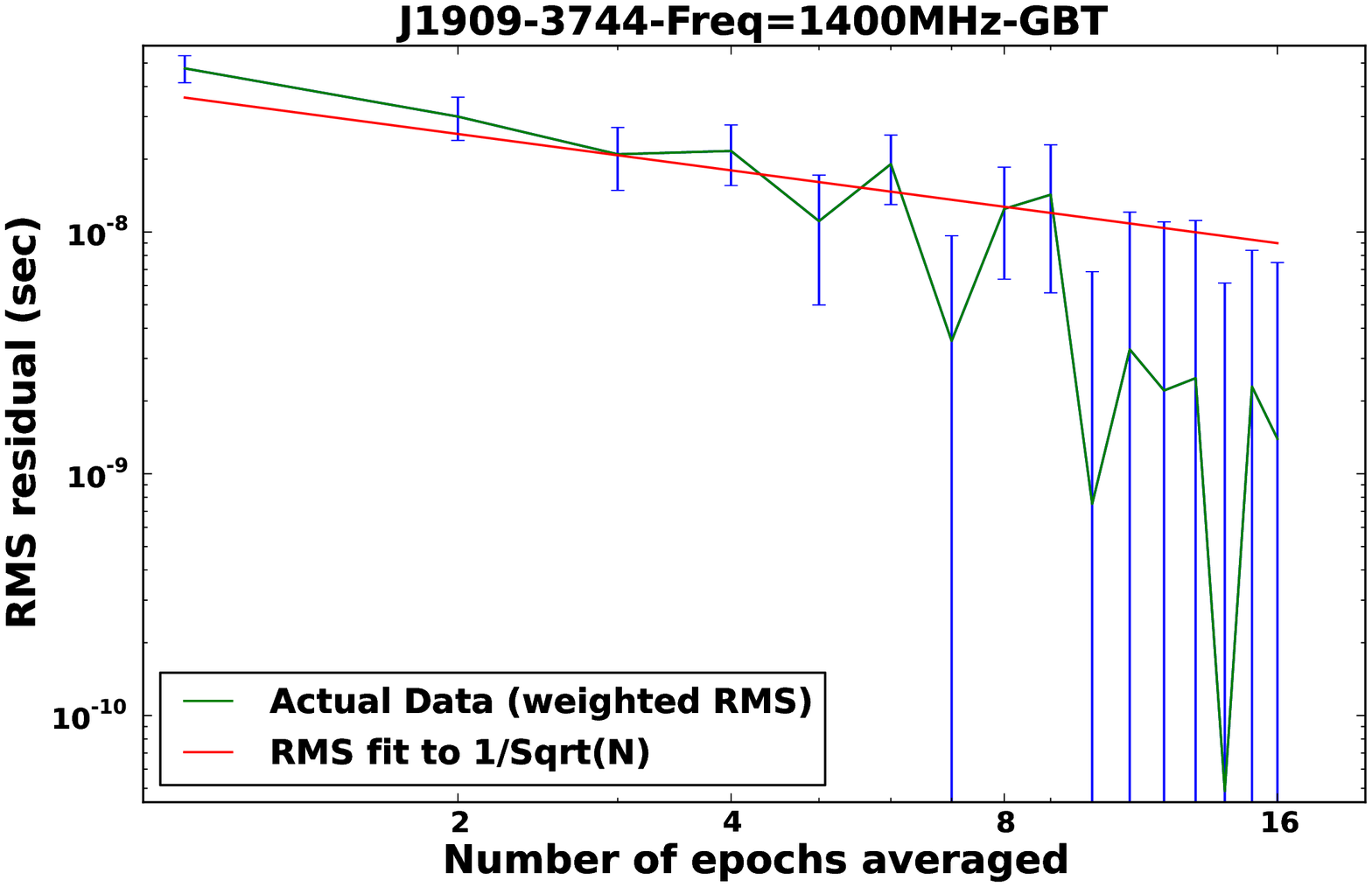}
                \caption{}
                \label{fig:residJ1909-3744b}
        \end{subfigure}
        \begin{subfigure}[b]{0.49\textwidth}
                \centering
                \includegraphics[width=\textwidth]{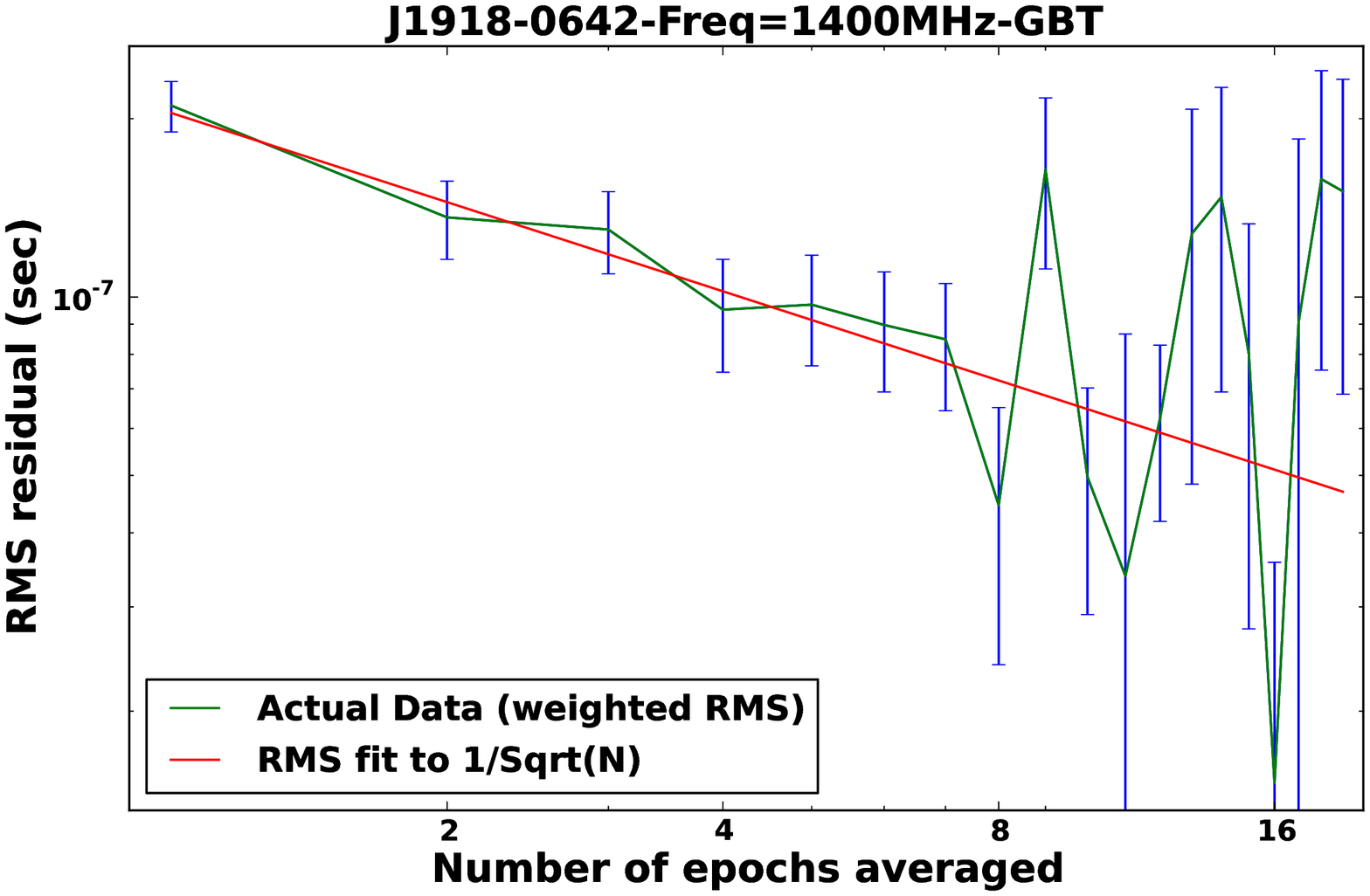}
                \caption{}
                \label{fig:residJ1918-0642b}
        \end{subfigure}%
        \begin{subfigure}[b]{0.49\textwidth}
                \centering
                \includegraphics[width=\textwidth]{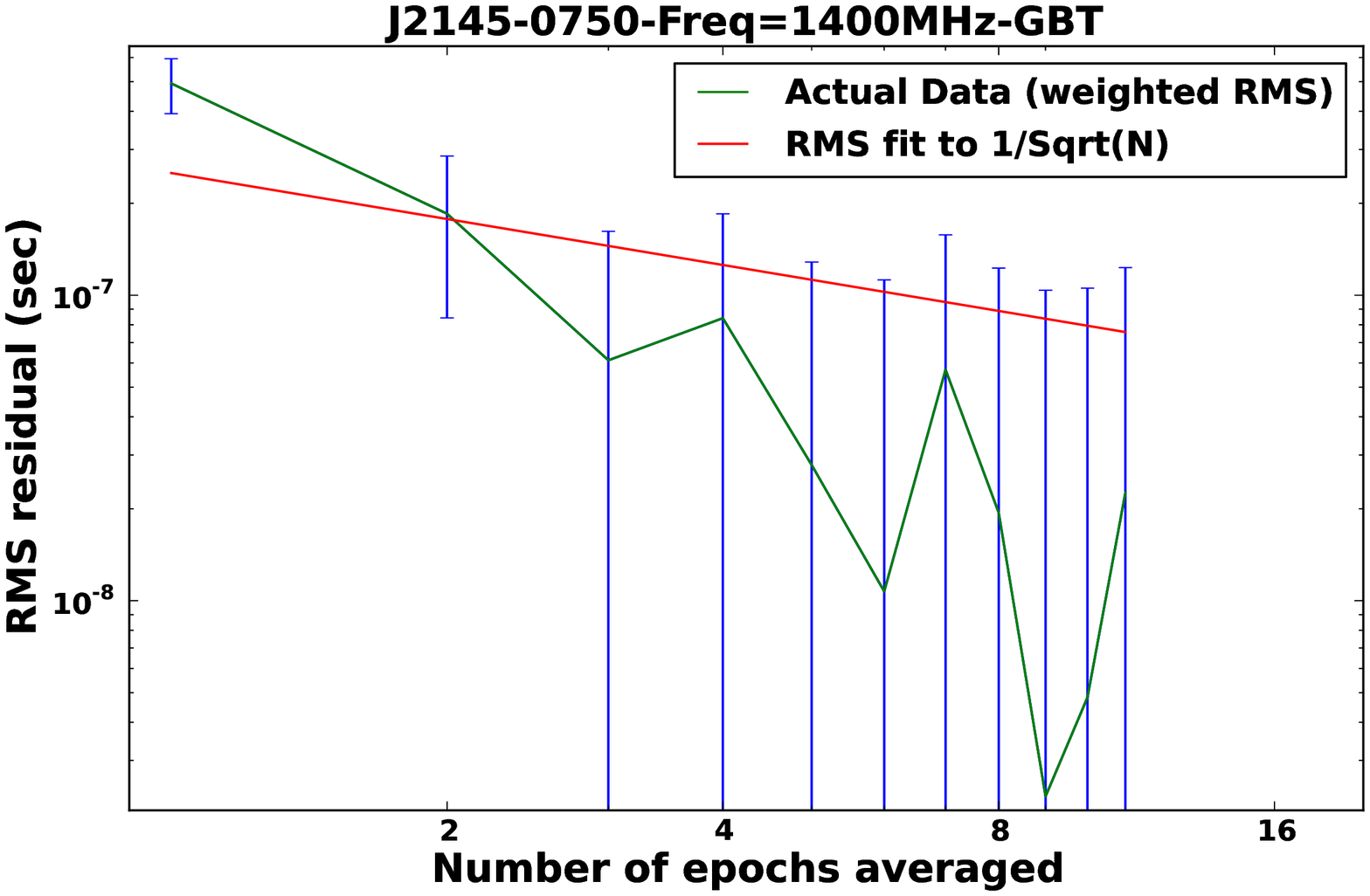}
                \caption{}
                \label{fig:residJ2145-0750b}
        \end{subfigure}
        \begin{subfigure}[b]{0.49\textwidth}
                \centering
                \includegraphics[width=\textwidth]{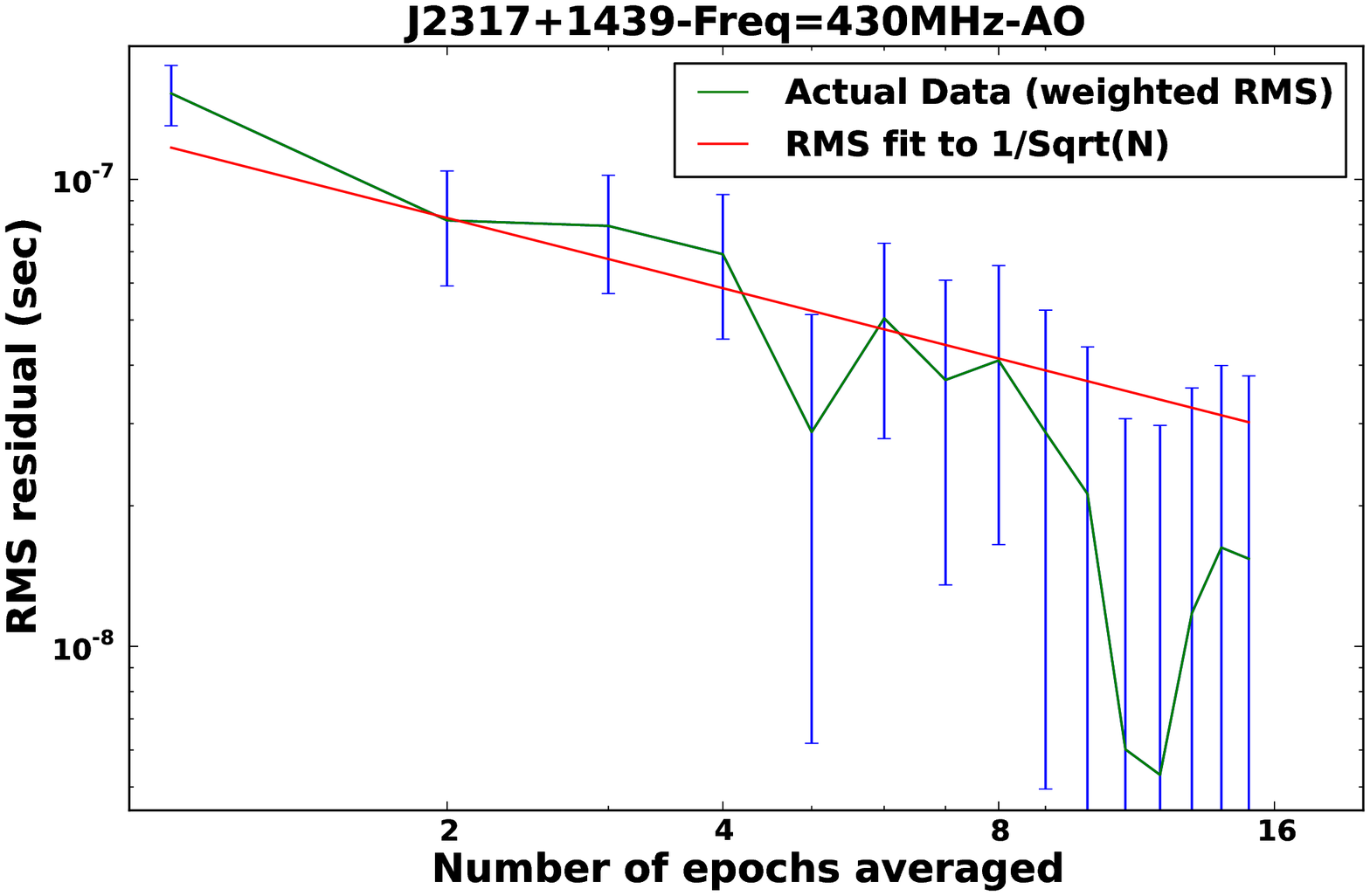}
                \caption{}
                \label{fig:residJ2317+1439b}
        \end{subfigure}
        \caption{RMS residual vs. the number of residuals averaged for the B1855+09
        when observed at 1410 MHz  (Panel a), PSR J1909$-$3744
        at 1400 MHz (Panel b), PSR J1918$-$0642 at 1400
          MHz (Panel c), PSR J2145$-$0750 at 1400 MHz (Panel d) and PSR J2317+1439 at 430 MHz (Panel e). 
	  The solid line represents the RMS fit to
          $\mathrm{N}^{-1/2}$.
          The plots show that within uncertainties, the
          RMS residual is proportional to $\mathrm{N}^{-1/2}$ for these
          pulsars.\label{fig:B1855+09J1909-3744J1918-0642J2145-0750J2317+1439} }
\end{figure}

 Figures \ref{fig:J0030+0451J0613-0200J1012+5307J1455-3330J1600-3053J17441134}, and \ref{fig:B1855+09J1909-3744J1918-0642J2145-0750J2317+1439}
are consistent with a WRMS proportional to N$^{-1/2}$ for PSRs J0030+0451, J0613$-$0200, J1012+5307, J1455$-$3330,
J1600$-$3053, J1744$-$1134, B1855+09, J1909$-$3744, J1918$-$0642, J2145$-$0750 and J2317+1439. 
The reduced $\chi^2$ values of the fit for each of the pulsars are presented in Table \ref{chi}. The pulsars just mentioned have 
reduced $\chi^2$ values close to 1 in the high frequencies, but the low frequencies give very small values. 
As described in \citet{Demorest13}, the low frequency band WRMS is expected to be suppressed because of the dispersion measure (DM) fitting, leading to small reduced $\chi^2$ values, see \S4. The best fit $\mathrm{WRMS}=\mathrm{aN}^{-1/2}$
line is shown in each Figures 1-9.

\begin{figure}[H]
        \centering
        \begin{subfigure}[b]{0.49\textwidth}
                \centering
                \includegraphics[width=\textwidth]{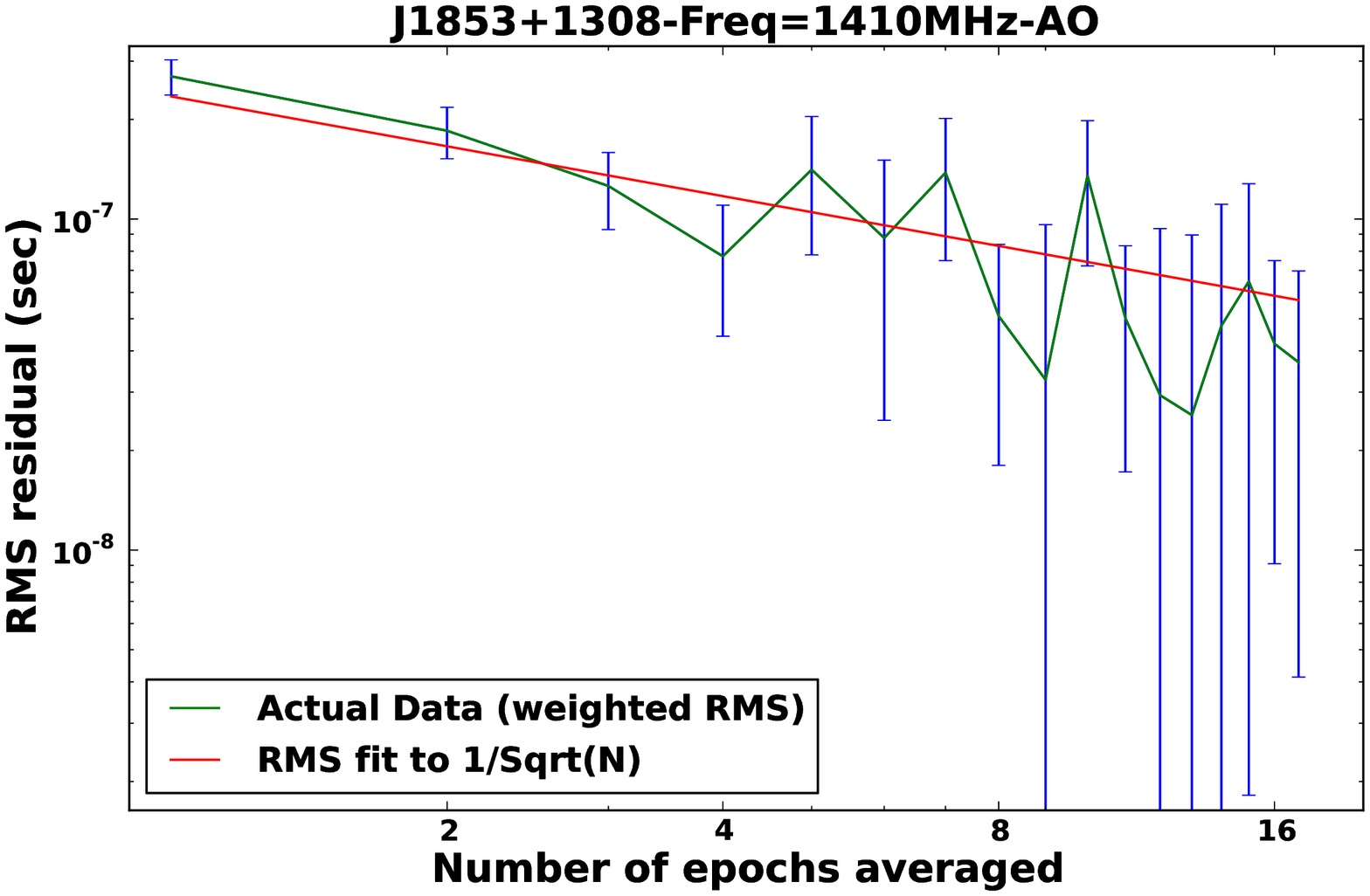}
                \caption{}
                \label{fig:residJ1853+1308}
        \end{subfigure}%
        \begin{subfigure}[b]{0.49\textwidth}
                \centering
                \includegraphics[width=\textwidth]{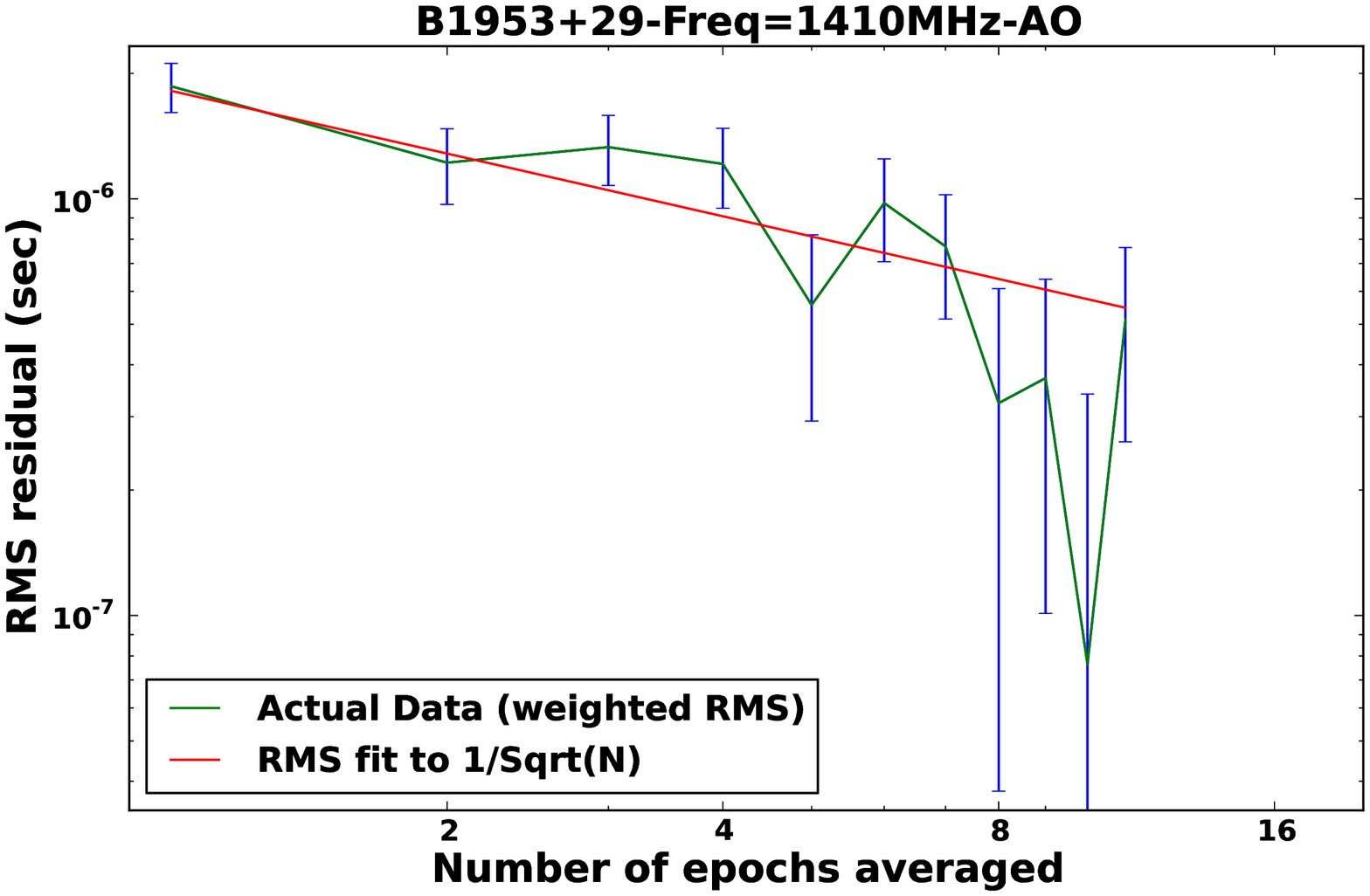}
                \caption{}
                \label{fig:residB1953+29}
        \end{subfigure}
        \begin{subfigure}[b]{0.49\textwidth}
                \centering
                \includegraphics[width=\textwidth]{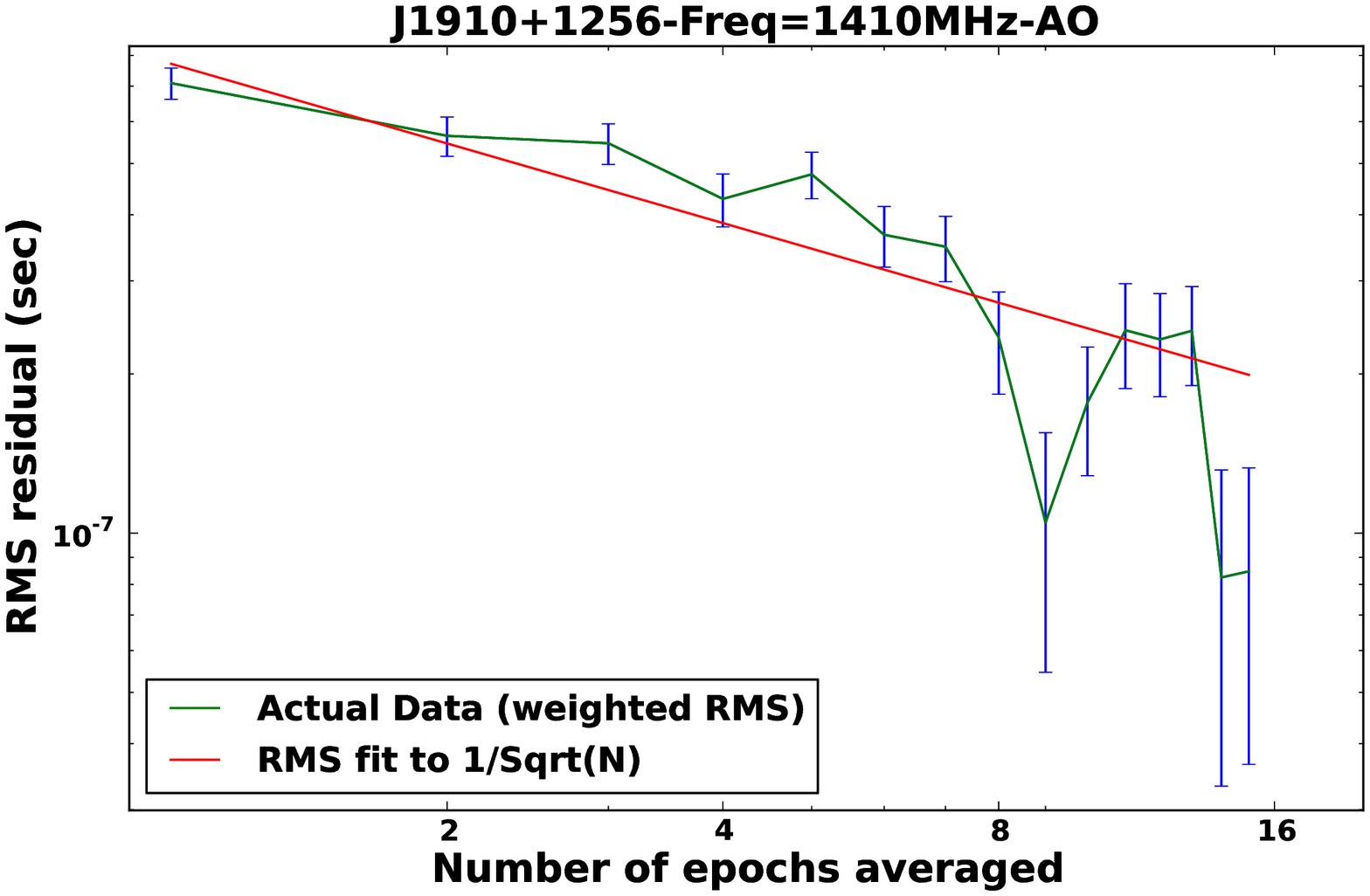}
                \caption{}
                \label{fig:cresidJ1910+1256}
        \end{subfigure}  
        \caption{RMS residual vs. the number of residuals averaged for PSR
          J1853+1308 when observed at 1410 MHz (Panel a), and PSR
          B1953+29 at 1410 MHz (Panel b) and PSR J1910+1256 at 1410
          MHz (Panel c). 
	  The solid line represents the RMS fit to
          $\mathrm{N}^{-1/2}$.
	The plots show that within uncertainties, the
          RMS residual is proportional to $\mathrm{N}^{-1/2}$ for these
          pulsars.}\label{fig:J1853+1308B1953+29J1910+1256}
\end{figure}

For PSRs B1953+29, J1853+1308 and J1910+1256 (Figure \ref{fig:J1853+1308B1953+29J1910+1256}), 
we only have one frequency band as mentioned previously. 
This limits the ability to correct for DM.  Despite this, the RMS still shows 
a general decreasing trend.

\begin{figure}[H]
        \centering
        \begin{subfigure}[b]{0.49\textwidth}
                \centering
                \includegraphics[width=\textwidth]{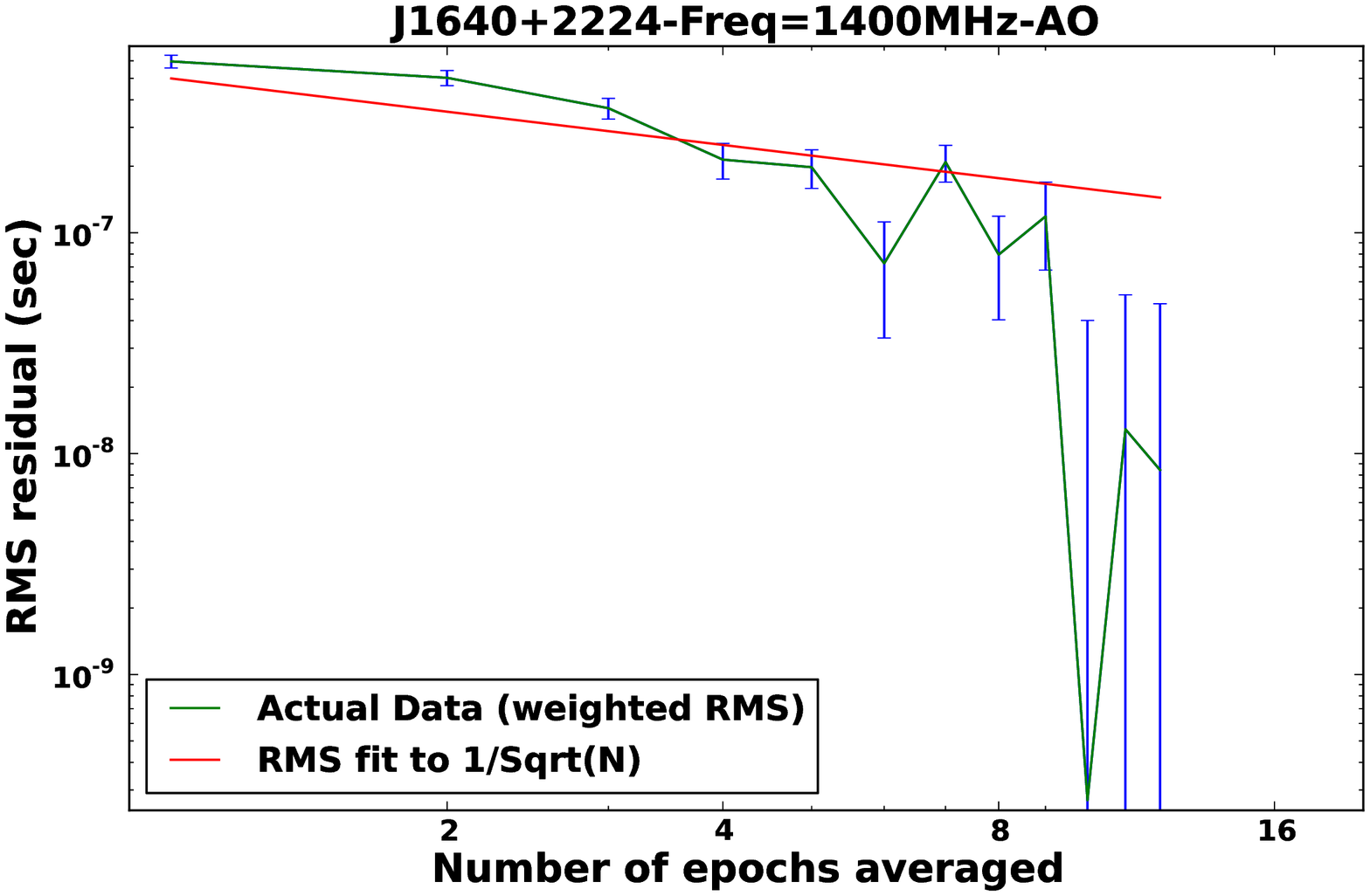}
                \caption{}
                \label{fig:residJ1640+2224b}
        \end{subfigure}%
        \begin{subfigure}[b]{0.49\textwidth}
                \centering
                \includegraphics[width=\textwidth]{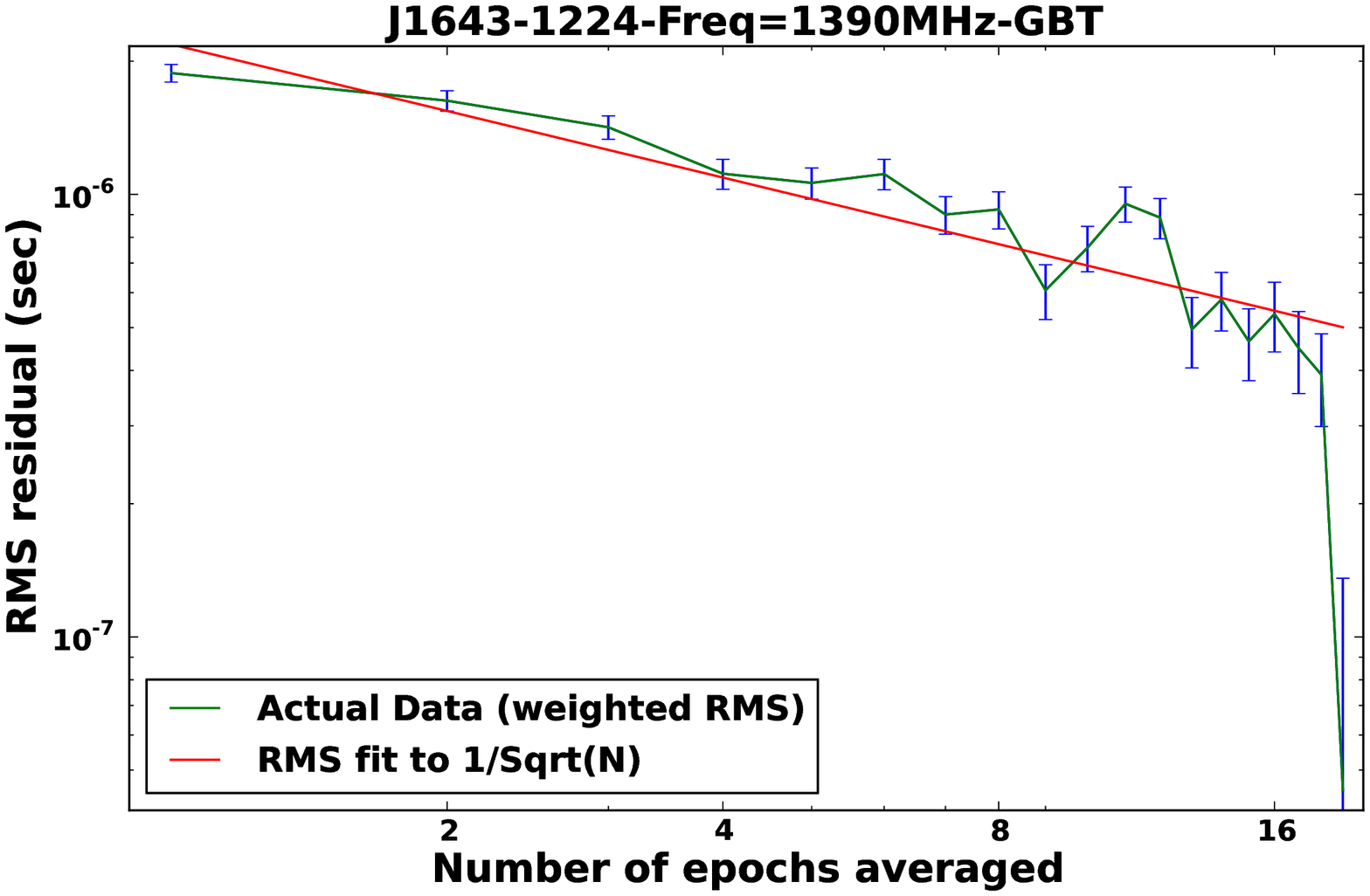}
                \caption{}
                \label{fig:residJ1643-1224b}
        \end{subfigure}
        \caption{RMS residual vs. the number of residuals averaged for the PSR J1640+2224
        when observed at 430 MHz (Panel a) and 1400 MHz (Panel b), and PSR J1643$-$1224
         at 830 MHz (Panel c) and 1390 MHz (Panel d).The solid line represents the RMS fit to
          $\mathrm{N}^{-1/2}$.
          The plots show that within uncertainties, the
          RMS residual is proportional to $\mathrm{N}^{-1/2}$ for these
          pulsars.\label{fig:J1640+2224J1643-1224} }
\end{figure}

There were only two pulsars that had a large reduced $\chi^2$, which is also evident 
when examining the plots 
in Figure \ref{fig:J1640+2224J1643-1224}. These were also pulsars identified by both 
\citet{Demorest13} and \citet{Perrodin15} as showing the largest evidence for red noise in the NANOGrav dataset.

\begin{figure}[H]
        \centering
        \begin{subfigure}[b]{0.49\textwidth}
                \centering
                \includegraphics[width=\textwidth]{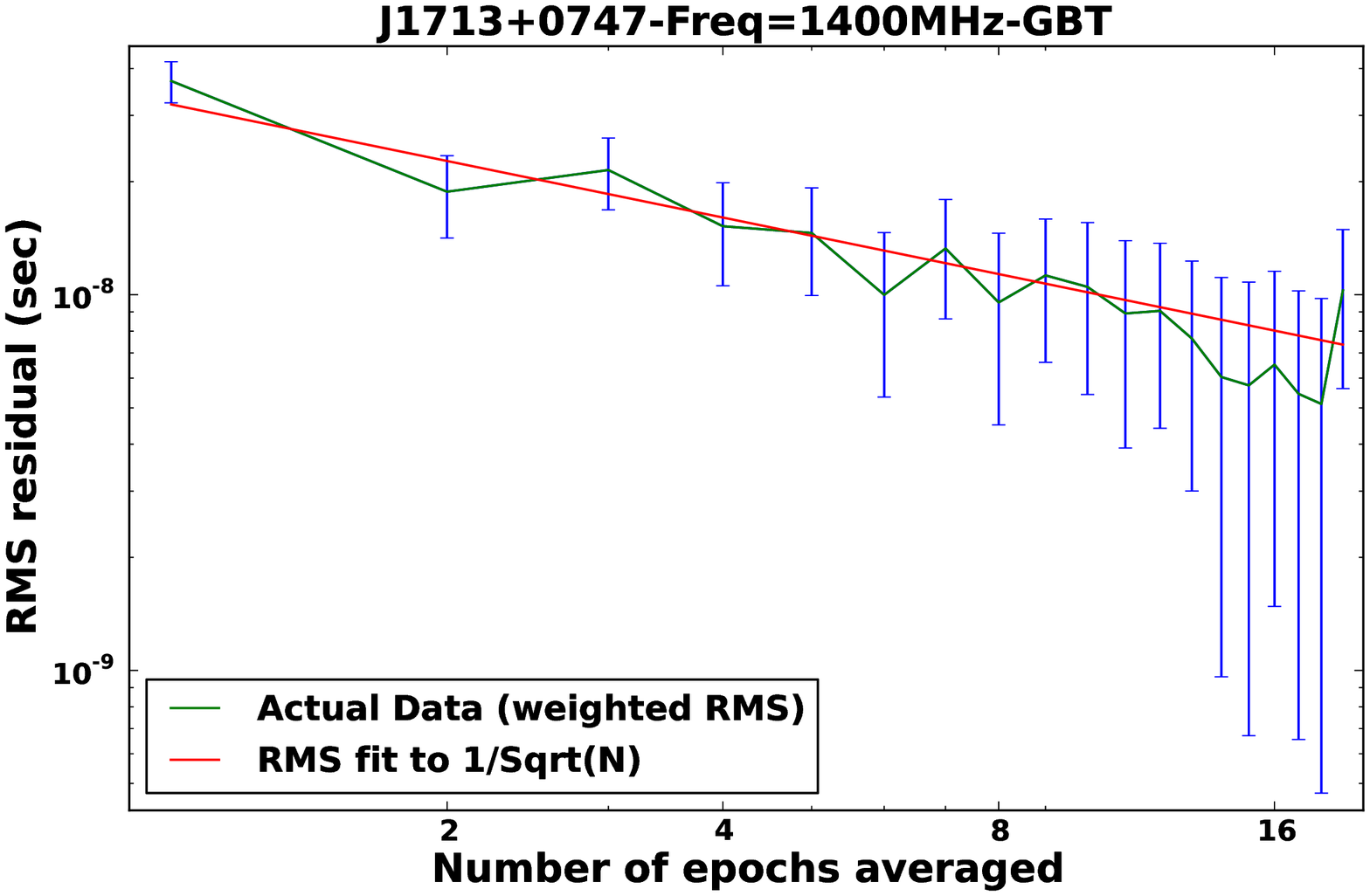}
                \caption{}
                \label{fig:residJ1713+0747c}
        \end{subfigure}%
        \begin{subfigure}[b]{0.49\textwidth}
                \centering
                \includegraphics[width=\textwidth]{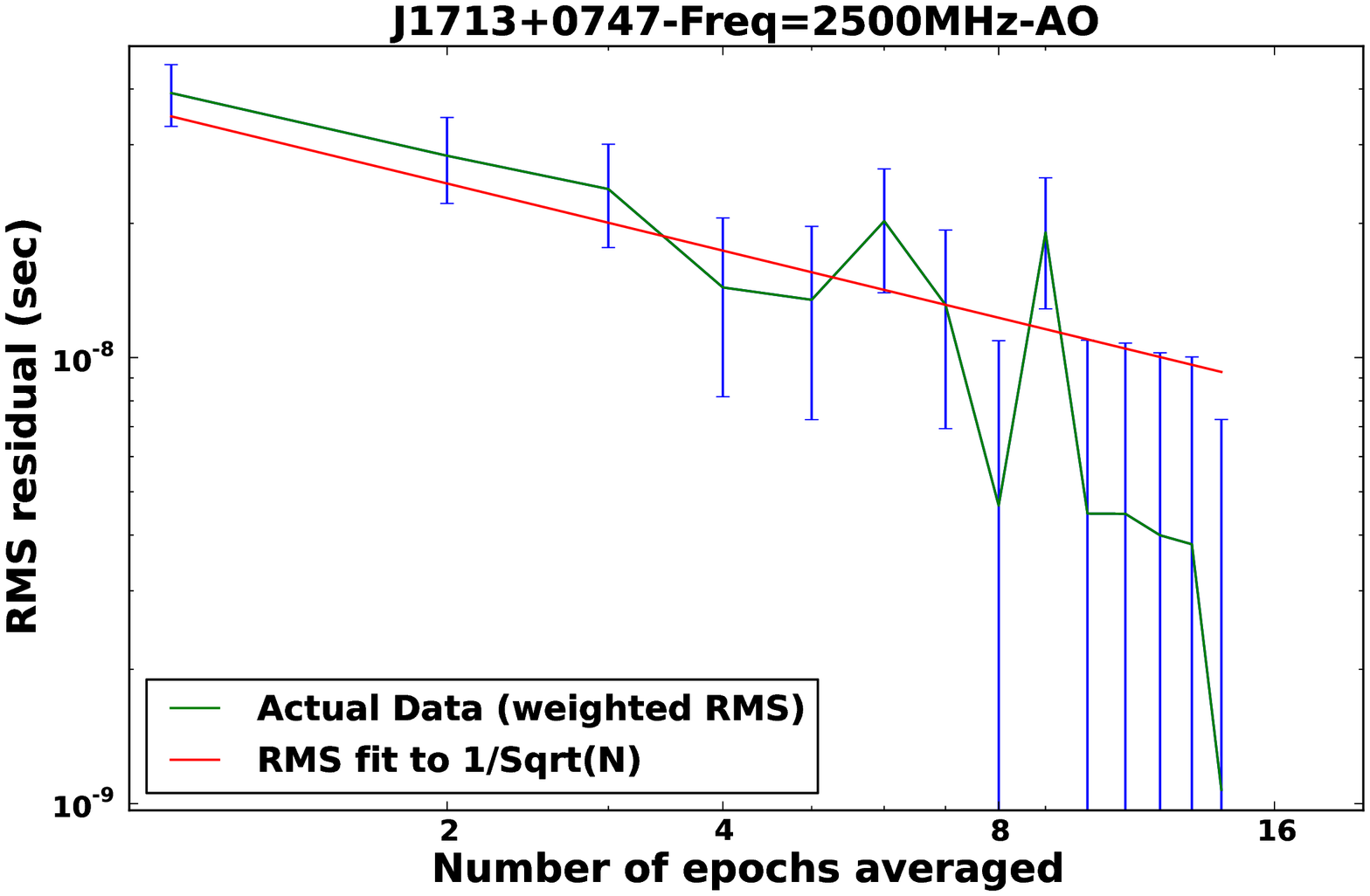}
                \caption{}
                \label{fig:residJ1713+0747d}
        \end{subfigure}    
        \caption{RMS residual vs. the number of residuals averaged for the
          pulsar when observed at 840 MHz (Panel a), at 1400 Mhz
          (Panel b), 1410 MHz (Panel c) and 2500 MHz (Panel d). 
	  The solid line represents the RMS fit to
          $\mathrm{N}^{-1/2}$. All plots show that within uncertainties, the
           RMS residual of PSR J1713+0747 follows the expected rate of decline of
          $\mathrm{N}^{-1/2}$.
          \label{fig:J1713+0747}
	  }
\end{figure}

PSR 1713+0747 (Figure \ref{fig:J1713+0747}) is unique in that it was observed at two different observatories. 
While each of the plots vary in shape, when looking at
the overall trends displayed, the RMS residual
decreases at the expected rate of $\mathrm{N}^{-1/2}$.. 

\begin{figure}[H]
  \centering
    \includegraphics[width=0.5\textwidth]{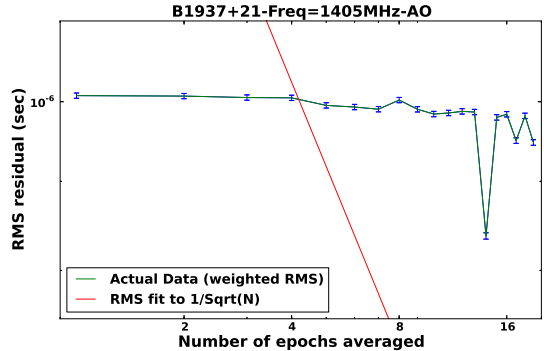}
  \caption{RMS residual vs. the number of residuals averaged for PSR B1937+21
    when observed at 1405MHz. The error bars on the data are too small to see.
    This data is from \citet{Lommen01t} and
    is not a part of the NANOGrav dataset. 
	  The solid line represents the RMS fit to
          $\mathrm{N}^{-1/2}$.
This plot shows, as expected for this pulsar with known red noise, 
that the RMS residual is not
    proportional to $\mathrm{N}^{-1/2}$.
	\label{fig:B1937+21}}
\end{figure}

For comparison, we analyzed data for PSR B1937+21 (Figure \ref{fig:B1937+21}) from
\citet{Lommen01t} in the same way. As a millisecond pulsar with a strong red noise signal, 
this provides an important test to ensure our method will demonstrate
 a noise floor in the presence of strong red noise.  We found the RMS does not decrease at all for this pulsar as 
more points are averaged. 

\section{Discussion}
The main point of this paper was to determine
 whether RMS decreases like $\mathrm{N}^{-1/2}$ when N, the number of adjacent observations 
 averaged, is increased.  ``N" is to be taken as a surrogate for the dwell time per 
 timing point, and we regard this as a test of whether increasing the dwell time 
 per point would reduce the residual RMS or not. This is a key question in planning future NANOGrav
 observations.
 
It is important to note the limitations 
of the test. It represents a 
necessary, but not sufficient condition, of white noise in the pulsar data. 
In other words, all white noise will pass the root N test, but not everything that passes the root N test is white noise.  For example, there may be a red noise process intrinsic to the pulsar, but the redness that such a process imprints on the timing residuals may be removed by the timing model fit.  Since determining the best timing model fit was beyond the scope of this work, we assumed the current NANOGrav fitting procedure of \cite{Demorest13}.
The residuals being analyzed have already passed through a timing model fit which 
has the property of `whitening' the data, especially in the flattening of low frequency power.  
Thus the noise processes themselves could be red while the timing residuals pass the root N test. 
If a pulsar fails this test, 
then we can conclusively state its spectrum is not white. However, displaying an RMS proporotional to root N does not necessarily mean the residuals themselves, nor the processes that created them, are true Gaussian processes.

Over-fitting is a concern for GW astronomy as it may eliminate a red noise signal from 
GWs by subsuming it into a fit. Indeed, this is a chief concern of \cite{Keith13} as the 
expected stochastic gravitational wave background (GWB) would be a red noise 
signal that may have significant power removed by an improper timing model fit, 
specifically in regards to fitting the time dependent nature of the DM.  Some recent analyses 
avoid this by fitting for the timing model parameters and GW signals simultaneously 
\citep{Ellis12, Arzoumanian14}, while others search the post-fit residuals while accounting 
for the effect of the timing model fit on their signal \citep{Jenet06,Christy15}.  
It is beyond the scope of this paper to study how an individual GW source will be affected by the
 \cite{Demorest13} timing model fit we assume.  However, we stress that even if this test does 
 not conclusively describe the underlying noise in a pulsar, decreasing the post-fit residual noise 
 will increase the sensitivity to a GW signal for any post-fit analysis and thus represents an 
 important aim of optimization studies.

We demonstrated that the residuals from 15 of the NANOGrav pulsars whose data we analyzed (B1855+09, B1953+29, J0030+0451, J0613$-$0200, J1012+5307, J1455$-$3330,
J1600$-$3053, J1713+0747, J1744$-$1134,
J1853+1308, J1909$-$3744, J1910+1256, J1918$-$0642, J2145$-$0750, and
J2317+1439) are consistent with the root N trend, with a few caveats summarized here. 

One striking feature of these results is that most pulsars in Figures 1,2 and 3 display a 
fit to root N that is better than the uncertainties would suggest, indicated by the 
reduced $\chi^2$ values being smaller than 1.
While it appears that most of the pulsars in NANOGrav have timing residuals that are consistent with
the $\mathrm{N}^{-1/2}$ trend, 
we caution the reader about this conclusion. There is some evidence that this 
pulsars are being overfitted by their models, however, they still follow the root N trend.

PSRs J2145$-$0750 and J2317+1439 do not have a lot of data. 
If we accumulate more data points for these pulsars, it is expected that these pulsars will continue
to be consistent with the trend of the RMS being proportional to N$^{-1/2}$.
On the other hand, PSR J1910+1256 has a lot of data, but since it was observed in only one 
frequency band, it could not be fitted for DM. The reduced $\chi^2$ value for this pulsar is 
large and it is evident that DM fitting is necessary to get an accurate understanding of the data.

There were only two pulsars that showed a deviation from the root N trend, namely J1640+2224, J1643$-$1224. This is evident by
their larger reduced $\chi^2$ values, and by their plots showing 
a shallower spectrum than the N$^{-1/2}$ fit. This could be an indication of a 
red noise component. This is a conclusion that was 
reached by both \citet{Demorest13} and \citet{Perrodin15}.

PSR B1937+21 is a known red pulsar (Figure \ref{fig:B1937+21}), so we 
include it to show how a truly red pulsar will behave in our scheme. 
The plot shows that there is no correlation between the RMS residual 
and $\mathrm{N}^{-1/2}$; this is also evident from its 
reduced $\chi^2$ value.
Many possible sources causing the
timing noise exhibited by this pulsar have been tested, but none have
been confirmed as the source of red noise \citep{Hobbs10a,
  Thorsett92, Shannon13, Shannon10}.  In any case, results from this pulsar are reassuring 
in that a pulsar dominated by red noise is not consistent with the root N test.

We wanted to determine 
whether increasing the dwell time on the NANOGrav pulsars will have
the expected effect of reducing the RMS residual in each pulsar.
In all cases the NANOGrav pulsars
follow the trend of
$\mathrm{RMS}\propto\mathrm{N}^{-1/2}$ steadily, at least until observed for
eight times as long ($N=8$) in each observing session. Using this conclusion, we expect to be able to
observe these 15 pulsars for up to 2-6 hours (depending on the pulsar's 
current dwell time of 15-45 minutes) in one observation, before we expect 
to see any inconsistency with RMS $\propto$ N$^{-1/2}$ relation. Indeed, this result has been shown 
for J1713+0747 in the recent, continuous 24 hour global campaign, 
where they observed a root T behavior \cite{Dolch14}.

Our claim that root N is a surrogate for root T 
assumes that the noise properties of the pulsars are stationary in time, 
which are discussed in \S2. Recent works by \citet{Ellis12} and references 
therein state that the NANOGrav data appears to display noise patterns that allow us to assume stationary statistics
with this data set.
If we can determine that the pulsar's timing residuals display noise that is 
consistent with the ``root T'' behavior, we can be confident that 
increasing the dwell time will yield lower noise in the timing residuals. 
This is critical input for decisions regarding 
future observations, specifically in determining how long to spend on each pulsar. 

\section{Conclusions}
The simplest way to decrease the noise in a PTA is to integrate over the 
noise for longer periods of time in each observation, e.g. go from 30-minute integrations on each pulsar 
to 1-hour integrations on each pulsar. However, this only works if the pulsar's timing residuals display 
noise that is consistent with the root N trend. We have averaged multiple (N) 
adjacent integrations, a substitute to actually increasing dwell time, to see whether the timing 
residuals display a root N behavior.  Our analysis 
on the NANOGrav pulsars shows that 15 of 17 are in this root N regime, 
with some displaying this all the way to an increase in dwell time by a factor of 8.

We note though that we only examined NANOGrav data, and the results 
we obtained are not necessarily the results received if this test was 
run on EPTA or PPTA data for these pulsars. We already see that just between
the two telescopes in NANOGrav we can get different results (PSR J1713+0747),
which may be different due to different DM removal algorithms \citep{Dolch14}.
 Improvements in our understanding will
come from attention to the dispersion removal algorithms. We saw evidence that sometimes
the current algorithm over-corrects the data. Further work will consider longer 
data sets where red noise may be more prominent. There are many other ways to characterize
the timing noise in pulsars \citep{Perrodin15, Ellis13, Cordes15}, but this test is a good first test
in determining the improvability of the timing noise of a pulsar and which pulsars should be included in a PTA.

Characterizing the timing noise in this way is only one aspect of optimizing
PTAs.  Optimization must be done with respect to a particular
goal; so is the goal of the PTAs to detect a stochastic background
first \citep{Lee12}, or a single source first \citep{Christy14}, or is it to fully characterize the
 GW sources we detect?  How we delegate telescope time will depend upon
the goal we choose in addition to the noise parameters. For
example, if the background is in fact stochastic, we may choose
different goals than if the background is dominated by a few bright
sources \citep{Babak12, Sesana08}. Knowing which pulsars' RMS can be improved is an important first step 
in optimizing a PTA.  In general, a lower RMS will produce more sensitive searches.  
The analysis done in this paper offers a measurement on 
the extent that the dwell time can be used to accomplish the goal of lowering the RMS.

\acknowledgments 
We would like to thank Ben Stappers and Gemma Janssen for some very helpful discussions and the hospitality of Jodrell
Bank Centre for Astrophysics while A.N.L. was on sabbatical. This project has been supported
by NSF AST CAREER 07-48580 and NSF PIRE 0968296.


\end{document}